\pdfoutput=1
\documentclass{article}

\usepackage{paper_conference,times}

\usepackage[utf8]{inputenc}
\usepackage[T1]{fontenc}
\usepackage{amsmath}
\usepackage{amssymb}
\usepackage{booktabs}
\usepackage{graphicx}
\usepackage{array}
\usepackage{textcomp}
\usepackage{caption}
\usepackage{subcaption}
\usepackage{placeins}
\usepackage{microtype}
\usepackage{url}
\usepackage[hidelinks]{hyperref}
 
\newif\iffinalversion
\finalversiontrue
\iffinalversion\iclrfinalcopy\fi

\newcommand{\plotdir}{plots}
\newcommand{\panelplot}[1]{%
  \includegraphics[width=\linewidth,height=0.55\textheight,keepaspectratio]{#1}}
\let\halfplot\panelplot
\newenvironment{chartfigure}{\begin{figure}[tb]}{\end{figure}}

\title{GVS5H: Zero-Shot Self-Orchestration \\
with Ledger-Based Control \\
Improves Coding in Language Models}

\author{%
  Victor Gao\thanks{Equal contribution.}\ \ \thanks{Persis Capital Inc.} \quad
  Vida Khosrowshahi\footnotemark[1]\ \ \footnotemark[2] \quad
  Ali Khosrowshahi\footnotemark[1]\ \ \footnotemark[2] \AND
  Xihao Sun\footnotemark[2] \quad
  Juhyun Lee \quad
  Ethan Tran\footnotemark[2] \quad
  Simon (Sang Won) Lee\footnotemark[2]\ \ \thanks{Corresponding author:
  slee@persisholdings.com}
}

\begin{document}

\maketitle

\lhead{}

\begin{abstract}
Frontier coding performance is typically attained with large, costly proprietary models. We introduce ledger-based zero-shot self-orchestration (GVS5H), a training-free method in which fresh instances of one model decompose problems and coordinate through a shared file system. 
Across eleven open and closed-weight models on the 100 latest hard LiveCodeBench problems,
the method yields as much as 25.6 points improvement, boosting several cheaper models to frontier-level performance. 
Orchestrated Qwen3.8 Flash Next scores 93.0\% against Fable 5's 90.4\% at 9\% of the cost, while the smaller Qwen3.8-27B reaches 92.4\%. 
Gains are not universal: some models are unchanged or worse. Transcript analysis attributes the gain to decomposition and persistent context. 
Inference-time organization can reach or exceed frontier coding accuracy at a fraction of the cost on self-hostable weights.
\end{abstract}

\section{Introduction}
\label{sec:intro}

Large language models (LLMs) are increasingly deployed not as single responses to a prompt. Instead, they are wrapped in a scaffold that performs multiple model calls, often interlaced with planning, tool calling, note taking,
and revision over multiple turns, on the intuition that decomposing a problem and making incremental progress should beat an attempt at answering everything in one pass.

In this paper, we test and quantify this intuition by measuring the performance of a single model call against a basic system consisting of multiple alternating orchestration and execution calls (manager and worker roles). Each call has a new context and communicates through a shared filesystem, with no access to additional tools.
 
We call the method studied here zero-shot self-orchestration (GVS5H: Gorilla vs.\ 5 Humans): the orchestrator is neither trained for orchestration nor
given task-specific demonstrations of how to decompose or coordinate the problem. Every model role is the same model in a fresh context
with a short generic prompt. This scaffold is further described in Section~\ref{sec:method}.

\subsection{Related work}
\label{sec:related}

Our test uses existing proprietary and open-weight language models, and what
we vary is the coordination layer above them. Prior systems differ mainly in how that layer is
decided. One line trains the coordinator: Fugu \citep{sakana2026} learns to assemble
and direct a pool of expert workers per query, and Conductor \citep{nielsen2025} is a 7B
model RL-trained to design worker communication topologies; earlier query routers learn to
pick a model per query \citep{ong2024}. A second fixes roles and hand-off order in advance,
so every problem flows through the same pipeline
\citep{hong2023,qian2023,li2023,huang2023,islam2024}, while frameworks such as AutoGen
\citep{wu2023} instead supply configurable conversable agents and leave the conversation
pattern open; Trinity \citep{xu2025} sits at the boundary with the first, fixing a
three-role workflow but learning which model fills each role. A third coordinates through a
shared store rather than a pipeline - ARIADNE \citep{wei2026} pairs a blackboard with MCTS
on program generation, LbMAS \citep{han2025} mediates all communication through a global
board - or decentralizes entirely, through debate \citep{du2023} or sampling and voting
\citep{li2024}; Mixture-of-Agents \citep{wangj2024a} instead aggregates in layers. A
separate line asks not how coordination is decided but whether reported gains survive
equalizing test-time compute \citep{wangj2024b,tran2026}.

Ours is the training-free counterpart to the first family and the dynamic counterpart to the
second. The manager is a fixed prompt with no learning, which lets us ask how much of the
orchestration benefit is available without training, and when it materializes. The list of tasks is dynamically generated and updated throughout the problem solving process rather than predetermined.
Its workspace is a shared store like ARIADNE's blackboard, but coordination is done by the manager
rather than through MCTS, reward models, or learned search policies. We do not attempt
an equal-token comparison across the same model - the loop necessarily spends more than a single call. 
Instead, we ask whether that spend buys better solutions, and whether it is more cost-effective than moving to a larger model.

\section{Method}
\label{sec:method}

Every role but the verifier is the same underlying model, called in a fresh context and
coordinating through a shared workspace on disk:

\begin{center}
\small
\begin{tabular}{@{}ll@{}}
\texttt{<ws>/task.md}     & the problem statement \\
\texttt{<ws>/plan.md}     & the manager's overarching plan \\
\texttt{<ws>/tasks.json}  & a debug mirror of the task list \\
\texttt{<ws>/notes.md}    & curated ideas / findings / partial proofs \\
\texttt{<ws>/solution.py} & the current solution \\
\end{tabular}
\end{center}

\begin{figure}[tb]
\centering
\includegraphics[width=\linewidth,height=0.46\textheight,keepaspectratio]%
  {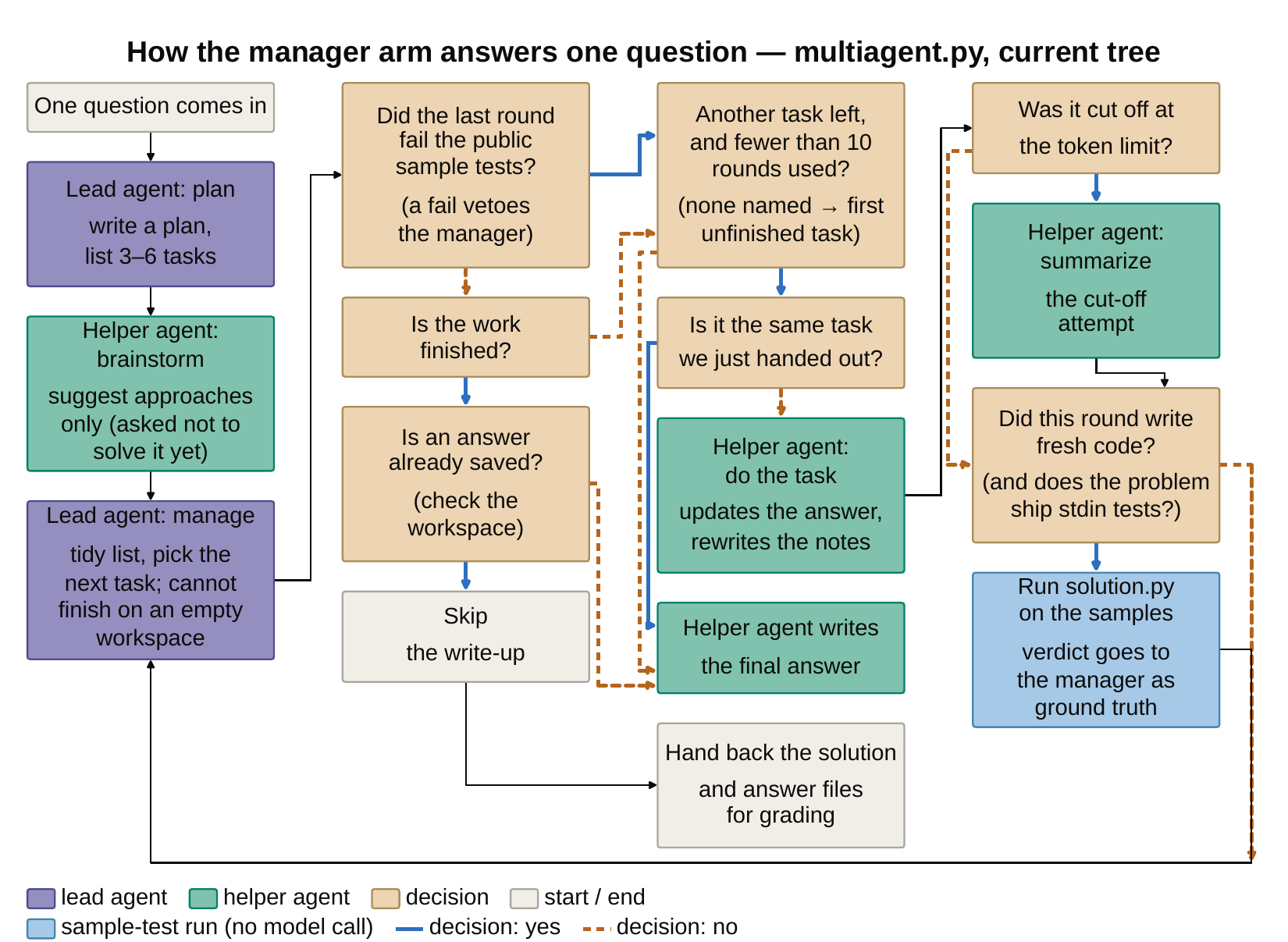}
\caption{\textbf{Manager--worker control flow}, as the v2 scaffold runs it. Boxes labelled
with an agent are model calls in a fresh context; the decisions, the start and end nodes,
and the sample-test run are harness logic and call no model. The boxes absent from the
original scaffold are the verifier of step 5 and the cut-off summarizer.}
\label{fig:loop}
\end{figure}

Workers never talk to each other; each sees the ledger plus the one task the manager
assigns. The control flow (Figure~\ref{fig:loop}) is:

\begin{enumerate}
  \item \textbf{Manager - plan.} The manager reads the problem and writes a 3--6 sentence
  strategy plus 3--6 concrete seed tasks.

  \item \textbf{Worker - brainstorm.} The first worker does not write a
  solution; it identifies the core difficulty, lists candidate approaches and pitfalls, and
  appends them to \texttt{notes.md}, proposing next steps.

  \item \textbf{Manager - manage.} The manager folds the plan and the brainstorm
  into one task list (merging duplicates, marking completed items, adding
  new sub-tasks), then either declares the problem done or names the single next task.

  \item \textbf{Worker - execute.} A fresh worker executes that one task and rewrites
  both \texttt{solution.py} and \texttt{notes.md} (the notes are replaced by a curated
  whole rather than appended to), then proposes remaining steps.

  \item \textbf{Verifier - run the public tests.} Whenever the round's worker produced a
  fresh candidate, the program is executed against the problem's public stdin tests where it
  has them. The pass/fail verdict, with the first failing case, is fed back to the manager
  and treated as ground truth: a failing run overrides a done verdict, forcing the loop to
  continue with a fix-or-switch task. Control returns to the manager (step 3).

  \item \textbf{Finalizer.} A finalization worker emits the definitive solution whenever the
  loop ends without a clean sign-off - the round budget is spent, the manager reissues a
  task it just handed out, or it names no task at all. The call is skipped when the manager
  declares the problem done and a usable solution is already on disk.
\end{enumerate}

Guards: a round budget of 10 manager$\to$worker cycles
(\texttt{MAX\_ITERS}); a no-progress guard (if the manager reissues the exact task it just
handed out, the loop stops); and a cut-off summarizer: a worker that hits the token cap
mid-attempt has its partial reasoning summarized by a fresh short call so its ideas still
reach the manager. The plan and notes files are capped when written, at 4,000 and 8,000
characters, so neither grows without bound across rounds.

\newpage
\paragraph{Baseline.} The single-call baseline is the same model given the same problem in exactly one call, with no
workspace, loops, or other roles. The single call receives the solver
system prompt verbatim; the manager arm's task workers receive that same prompt wrapped in a
subagent preamble, and their user message additionally
carries the plan, the current notes and the current solution. The two arms differ only in
that scaffold. Section~\ref{sec:setup} gives the per-role temperatures.

\subsection{The original scaffold}
\label{sec:scaffold-v1}

The loop above is v2, which produced the pinned-backend results and is drawn in
Figure~\ref{fig:loop}. The original scaffold produced the OpenRouter-served set of
Section~\ref{sec:openrouter}. It runs the same loop on closely matched prompts, with five
differences, all of them on the manager arm: it allows 4 manager$\to$worker cycles rather than 10
(\texttt{MAX\_ITERS}), it has no public-test verifier at step 5, it has no cut-off summarizer, its
workers append to \texttt{notes.md} rather than rewriting it, and the plan and notes files feeding
a prompt are unbounded rather than capped.

\subsection{Experimental setup}
\label{sec:setup}

\paragraph{Benchmark.} LiveCodeBench code-generation, release\_v6 \citep{jain2024}: the 100
most recent problems in the hard split by contest date, which limits training-data
contamination \citep{jain2024}. We refer to this set as LCB-100. Solutions are graded by
LiveCodeBench's own hidden-test evaluator, after the two corrections described in
Appendix~\ref{app:evaluator}; every number below is post-correction. The verifier of step 5
runs only stdin-format public tests, which 73 of the 100 problems carry; the other 27 are
LeetCode-style, with call-based public tests the engine does not run, so those get no verifier
signal and are checked by the hidden-test grader alone.

\paragraph{Models and serving.} Eleven model configurations: Qwen3.5-9B \citep{qwen2026a} (9B),
Qwen3.6-35B-A3B \citep{qwen2026b} (35B total, 3B active per token), Qwen3.8-27B
\citep{qwen2026c} (27B, open weights, served locally in FP8), Qwen3.8-Flash-Next
\citep{qwen2026d} (125B total, 6B active per token, open weights, served locally in FP8),
DeepSeek-V4.1-Flash \citep{deepseek2026} (552B total, 16B active per token at decode, open
weights, served by the first-party DeepSeek API), MiniMax-M3 \citep{minimax2026}
(428B total, ${\sim}23$B active), Kimi-K3 \citep{moonshot2026} (${\sim}2.8$T total,
${\sim}16$ of 896 experts active per token), and the closed frontier models Opus-5
\citep{anthropic2026}, the two GPT-5.6 variants Terra and Luna \citep{openai2026}, and Claude
Fable 5 \citep{anthropic2026}, all four of undisclosed size. Serving splits the set in two.
The six pinned-backend models are each served on one fixed backend for v2 comparison, because
gateway routing proved the dominant source of run-to-run noise (Section~\ref{sec:discussion}):
Terra and Luna by the OpenAI API, Qwen3.8-27B and Qwen3.8-Flash-Next by our own vLLM,
DeepSeek-V4.1-Flash by the first-party DeepSeek API, and Fable 5 by the
Anthropic Messages API. The five OpenRouter-served models ran through a
single OpenAI-compatible gateway on the original scaffold, and are reported separately in
Section~\ref{sec:openrouter} because both the serving path and the scaffold version differ.
Fable 5 was run single-only, so it contributes no manager delta.

\paragraph{Temperatures.} The manager arm's generative calls use slightly higher temperatures,
where the work is generative rather than executive: 0.3 to write the plan and 0.4 to brainstorm,
against 0.2 for task execution and for curating the task list. The single-call baseline runs at
0.2, matching the workers. These settings are identical in both scaffold versions.

\paragraph{Conditions.} The pinned-backend set runs at one setting (128k output cap,
reasoning on, $\times 5$ passes) with the five models in both arms. The cap is
a per-call \texttt{max\_tokens} bound, not a context-window limit. The OpenRouter-served set
runs at three: 128k reasoning-on $\times 1$; 16k reasoning-off $\times 5$; and 128k
reasoning-off $\times 1$ as a control isolating the small cap. Appendix~\ref{app:protocol}
gives the reasoning-effort settings per provider, the instrumentation, and the cap-matching
replay used for Qwen3.8-27B's single arm.

\paragraph{Statistics.} Reported $\pm$SD is the sample SD (divisor $n-1 = 4$) of the five
pass scores. Five-pass comparisons use a paired sign-flip permutation test on per-problem
mean $\Delta$ ($n = 100$ problems; $2\times10^{5}$ resamples), Holm-corrected where a family
of models is tested; 95\% $t$-intervals for mean pass@1 across the five passes
($\mathrm{df} = 4$). Pooled problem$\times$pass agreement uses exact McNemar on the
discordants, treating cells as independent.

\section{Results}
\label{sec:results}

\subsection{Pinned-backend arms, five passes}
\label{sec:pinned}

\begin{chartfigure}
\centering
\halfplot{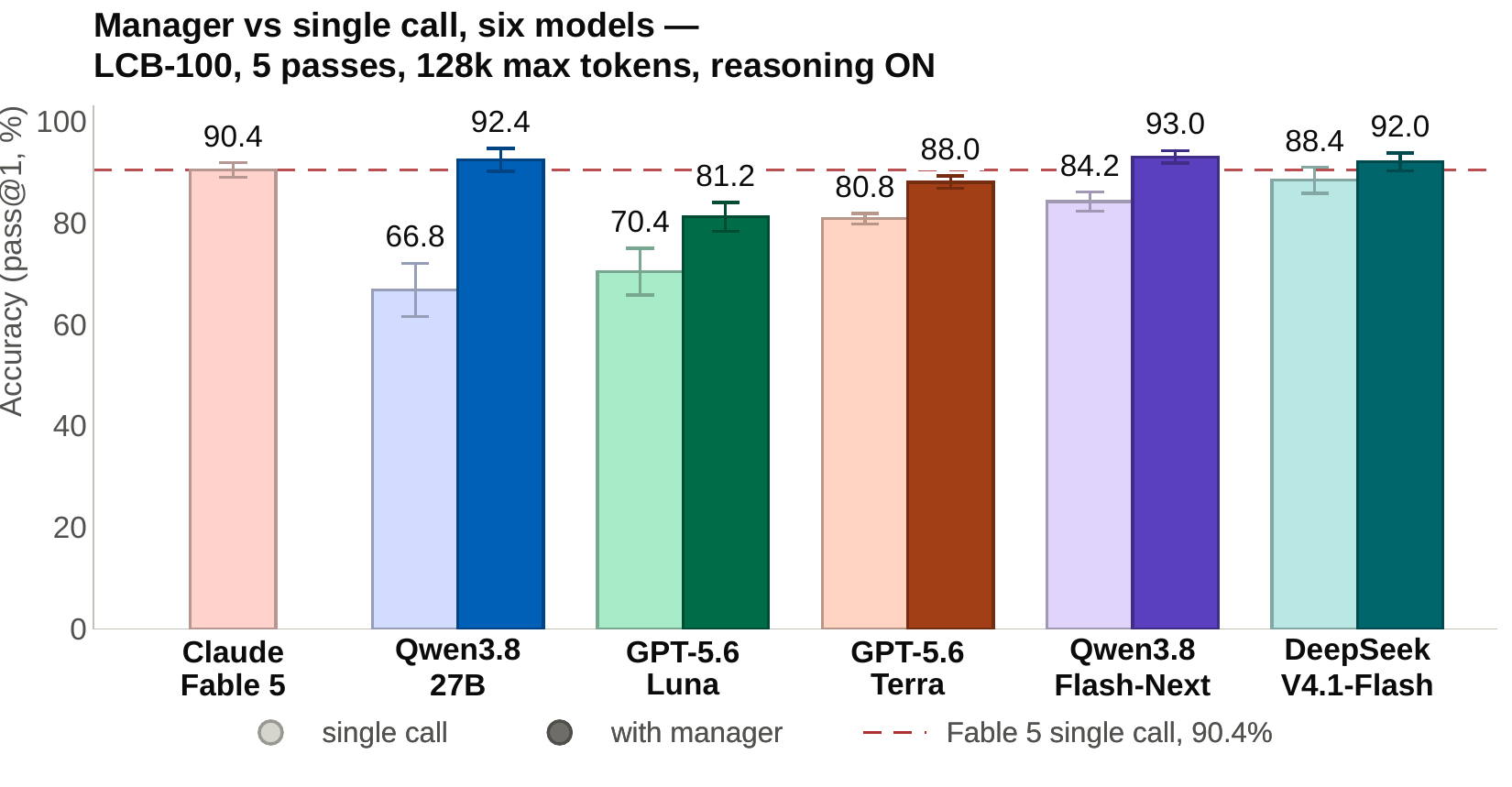}
\caption{\textbf{Manager vs.\ single call.} The scores from Table~\ref{tab:main} in a graph. Bars are pass@1 on the same 100 problems, the line through each the 95\% CI across the 5 passes (t, df = 4). The left bar of each pair is the single call, light; the manager is beside it, dark; colour is the model. Fable 5 ran single-only, so it has one bar, which the dashed rule carries across the chart.}
\label{fig:main}
\end{chartfigure}

\begin{table}[!hb]
\centering
\small
\caption{\textbf{Manager vs.\ single call, five passes.} pass@1 on LCB-100 at a 128k cap, reasoning on; $\pm$ is a 95\% $t$ interval across the five passes (df = 4). $\Delta$ is in percentage points, against the model's own single call and against Fable 5's single call. Each $p$ is a paired sign-flip permutation test, unit = problem ($n = 100$), Holm-corrected within its family of five; $<$ marks the permutation floor. \textbf{95\% bound} is the one-sided lower bound on $\Delta$ vs Fable 5 over the same per-problem differences (t, df = 99): the largest deficit the data leave open.}
\label{tab:main}
\begin{tabular}{@{}l@{\hspace{0.55em}}c@{\hspace{0.55em}}c@{\hspace{0.55em}}r@{\hspace{0.55em}}c@{\hspace{0.55em}}r@{\hspace{0.55em}}r@{\hspace{0.55em}}c@{}}
\toprule
\textbf{Model} & \textbf{\shortstack{Single\\call}} & \textbf{\shortstack{With\\manager}} & \textbf{\shortstack{$\Delta$ vs\\single}} & \textbf{$p$} & \textbf{\shortstack{$\Delta$ vs\\Fable 5}} & \textbf{\shortstack{95\%\\bound}} & \textbf{$p$} \\
\midrule
Claude Fable 5 & $90.4 \pm 1.4$ & --- & --- & --- & reference & --- & --- \\
GPT-5.6-Terra & $80.8 \pm 1.0$ & $88.0 \pm 1.2$ & $+7.2$ & $8.0\times10^{-5}$ & $-2.4$ & $-5.8$ & $0.88$ \\
GPT-5.6-Luna & $70.4 \pm 4.6$ & $81.2 \pm 2.8$ & $+10.8$ & $<2.5\times10^{-5}$ & $-9.2$ & $-13.4$ & $0.0013$ \\
Qwen3.8-27B & $66.8 \pm 5.2$ & $92.4 \pm 2.3$ & $+25.6$ & $<2.5\times10^{-5}$ & $+2.0$ & $-0.8$ & $0.88$ \\
Qwen3.8-Flash-Next & $84.2 \pm 1.8$ & $93.0 \pm 1.2$ & $+8.8$ & $<2.5\times10^{-5}$ & $+2.6$ & $-0.1$ & $0.56$ \\
DeepSeek-V4.1-Flash & $88.4 \pm 2.6$ & $92.0 \pm 1.8$ & $+3.6$ & $0.002$ & $+1.6$ & $-1.3$ & $0.88$ \\
\bottomrule
\end{tabular}
\end{table}

Our headline condition is six models, each run five independent times on LCB-100 at a 128k
cap with reasoning on. Every arm is served on a pinned backend and runs the v2 scaffold except Fable 5. The
earlier runs of the OpenRouter-served set (Section~\ref{sec:openrouter}) cover more models,
but only their 16k reasoning-off condition repeats; the rest are one pass each, through a
gateway whose routing proved the main source of run-to-run noise.

\textbf{All five models with both arms benefit from the manager.} It also narrows
the wide run-to-run spread for Luna and Qwen3.8-27B. Qwen3.8-27B gains most, its manager arm
reaching Fable 5's single-call range here, and Opus-5's in the OpenRouter-served set. The
direction matches that set's reasoning-on pattern: the smaller the model, the more the
scaffold has to add.

\subsection{What the scaffold costs} 
\label{sec:cost}

The manager uses more tokens by replacing one call
with a plan, a brainstorm, up to ten worker rounds and a finalization pass. On average, the scaffold roughly triples the bill. Every arm
priced here is at the same 128k output cap, Qwen3.8-27B's single call included
(Appendix~\ref{app:protocol}).

\begin{table}[!t]
\centering
\small
\caption{\textbf{What one pass cost each arm.} List rate $\times$ the tokens it consumed. The two $p$ columns test each manager arm against its own single call and against Fable 5's single call: Welch over the five passes, Holm-corrected across 11 comparisons. Qwen3.8-27B's single arm is the 128k cap-matched one; at 250k it costs \$26.95/pass. Retried and discarded attempts are counted: they were generated and would be billed. The cheapest arm is GPT-5.6-Luna single at \$0.41 a pass and the most accurate is Qwen3.8-Flash-Next manager at \$5.76 --- a 14$\times$ spread in price for +22.6 points.}
\label{tab:cost}
\begin{tabular}{@{}l@{\hspace{0.5em}}l@{\hspace{0.5em}}r@{\hspace{0.5em}}r@{\hspace{0.5em}}r@{\hspace{0.5em}}r@{\hspace{0.5em}}c@{\hspace{0.5em}}c@{}}
\toprule
\textbf{Arm} & \textbf{\shortstack{Rate \$/MTok\\in / out}} & \textbf{In (MTok)} & \textbf{Out (MTok)} & \textbf{\$/pass} & \textbf{\$/solved} & \textbf{\shortstack{$p$ vs\\single}} & \textbf{\shortstack{$p$ vs\\Fable 5}} \\
\midrule
\shortstack[l]{Qwen3.8-27B\\single} & \$0.21 / \$2.55 & 0.0753 & 7.4247 & \$18.95 & \$0.28 & --- & --- \\
\shortstack[l]{Qwen3.8-27B\\manager} & \$0.21 / \$2.55 & 1.5053 & 18.6277 & \$47.82 & \$0.52 & $1.5\times10^{-4}$ & $6.1\times10^{-4}$ \\
\shortstack[l]{Qwen3.8-Flash-Next\\single} & \$0.15 / \$0.47 & 0.0753 & 5.7751 & \$2.73 & \$0.032 & --- & --- \\
\shortstack[l]{Qwen3.8-Flash-Next\\manager} & \$0.15 / \$0.47 & 1.2930 & 11.8380 & \$5.76 & \$0.062 & $3.5\times10^{-4}$ & $3.6\times10^{-7}$ \\
\shortstack[l]{DeepSeek-V4.1-Flash\\single} & \$0.30 / \$1.20 & 0.0680 & 2.8626 & \$3.46 & \$0.039 & --- & --- \\
\shortstack[l]{DeepSeek-V4.1-Flash\\manager} & \$0.30 / \$1.20 & 1.4993 & 12.4873 & \$15.43 & \$0.17 & $1.6\times10^{-4}$ & $2.2\times10^{-9}$ \\
\shortstack[l]{GPT-5.6-Luna\\single} & \$0.20 / \$1.20 & 0.0661 & 0.3299 & \$0.41 & \$0.006 & --- & --- \\
\shortstack[l]{GPT-5.6-Luna\\manager} & \$0.20 / \$1.20 & 1.1686 & 1.0522 & \$1.50 & \$0.018 & $6.4\times10^{-5}$ & $1.3\times10^{-6}$ \\
\shortstack[l]{GPT-5.6-Terra\\single} & \$2 / \$12 & 0.0661 & 0.2728 & \$3.41 & \$0.042 & --- & --- \\
\shortstack[l]{GPT-5.6-Terra\\manager} & \$2 / \$12 & 1.1098 & 0.7911 & \$11.71 & \$0.13 & $1.6\times10^{-4}$ & $1.9\times10^{-8}$ \\
\shortstack[l]{Fable 5\\single} & \$10 / \$50 & 0.0899 & 1.2043 & \$61.11 & \$0.68 & --- & reference \\
\bottomrule
\end{tabular}
\par\vspace{0.6ex}
{\footnotesize List rates: Qwen3.8-27B \citep{openrouter2026}, Qwen3.8-Flash-Next \citep{openrouter2026flash}, GPT-5.6-Luna and GPT-5.6-Terra \citep{openai2026price}, DeepSeek-V4.1-Flash \citep{deepseek2026price}, Fable~5 \citep{anthropic2026price}. OpenAI uses short-context rates and Anthropic and DeepSeek use the base rates; no long-context tier, batch/off-peak discount, or prompt-caching multiplier is applied. Qwen and DeepSeek rates were read on 2026-09-15.}
\end{table}

\begin{figure}[htbp]
\centering
\panelplot{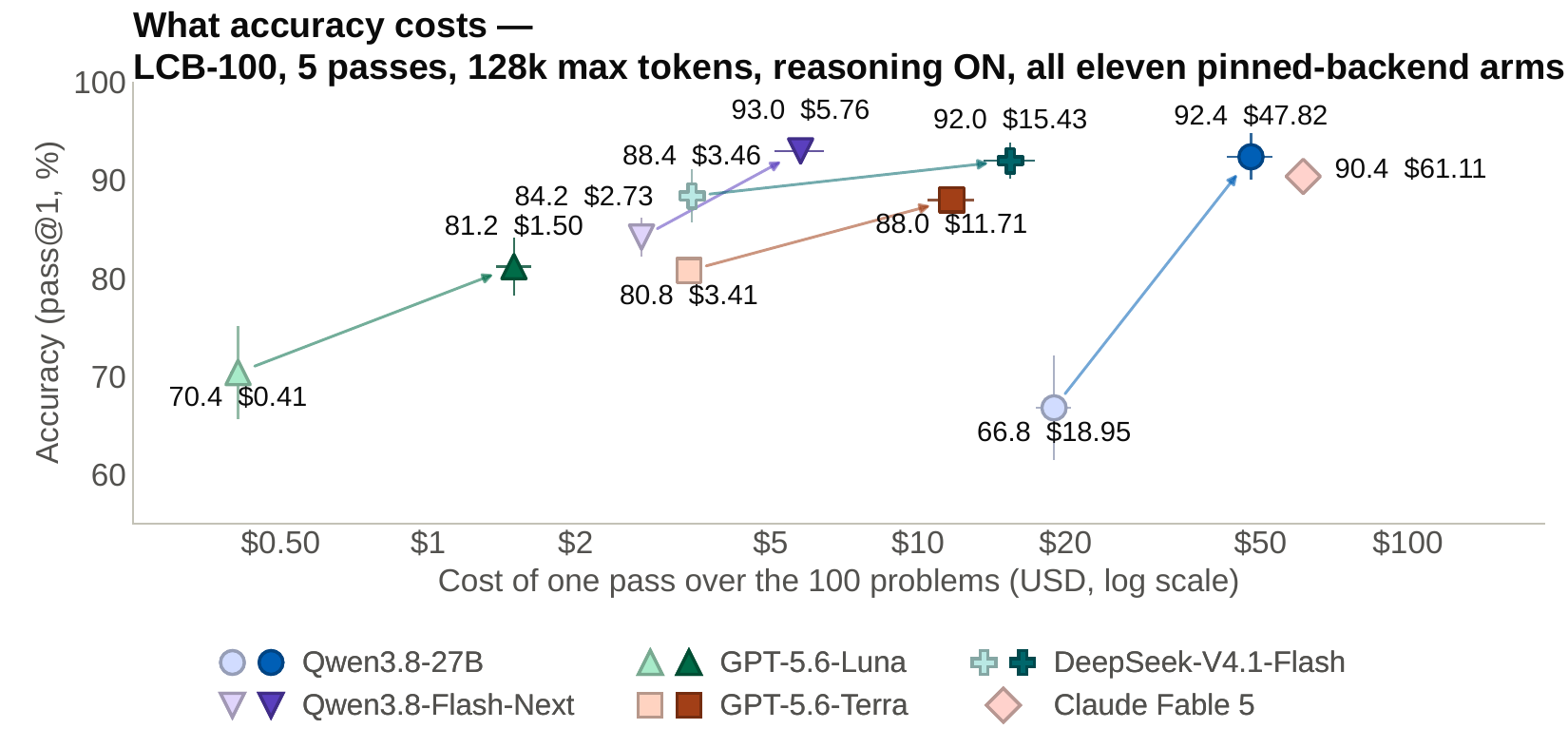}
\caption{\textbf{Cost against accuracy.} One point per arm; an arrow runs from the single call to the manager of each model that has both. x is dollars for one pass over the 100 problems (log scale), y is pass@1; the bar through each point is the 95\% CI across the 5 passes (t, df = 4). Light fill = single call, dark = with manager; marker shape is the model.}
\label{fig:cost-vs-score}
\end{figure}

\textbf{The scaffold improves performance for a higher price.}
Figure~\ref{fig:cost-vs-score} plots price against accuracy: every model's manager arm gains accuracy over its single
call, and costs more. On average, the number of output tokens increases to about 3x baseline.
 
\textbf{The base cost varies across models.} A single-call pass on Qwen3.8-27B emits more
than twenty times the output tokens of either OpenAI model. The total cost depends on the model's reasoning tendencies in addition to its per-token rate and scaffold.

\paragraph{Comparisons along the frontier.}
In many cases, the manager arm of a cheaper model can have a lower cost than a single call to an expensive model, while providing a higher accuracy. Most notably, Qwen3.8-Flash-Next scores at or above Fable 5 for a tenth of the price. Qwen3.8-27B also edges ahead of Fable 5 for a slightly lower price, with a much lower hardware requirement to host locally. Within the GPT-5.6 family, GPT-5.6-Luna with a manager matches GPT-5.6-Terra’s single call on accuracy at under half the price, making a clear case for spending on the scaffold rather than the model.

\FloatBarrier
\subsection{The effect of truncation}
\label{sec:truncation}

Each model generation is cut off once its thinking and output exceed the token limit. In the manager case, a new, fresh round is started. In the single call run or the final call of a manager run, if no valid code was generated before the cutoff, the task is marked as a fail. Table~\ref{tab:caps} shows the frequency of these truncations.

\begin{table}[htbp]
\centering
\caption{\textbf{Cap hits and empty solutions per arm.} A \emph{cap hit} is a generation that stopped at the token limit (\texttt{finish\_reason=length}).  \emph{No code} counts  any run containing no code, due to a cap hit or other reason.}
\label{tab:caps}
\small
\begin{tabular}{@{}l@{\hspace{0.45em}}c@{\hspace{0.45em}}c@{\hspace{0.45em}}c@{\hspace{0.45em}}c@{\hspace{0.45em}}c@{\hspace{0.45em}}c@{\hspace{0.45em}}c@{}}
\toprule
\textbf{Model} & \textbf{\shortstack{Cap hits\\single}} &
\textbf{\shortstack{Cap hits\\manager}} & \textbf{\shortstack{No code\\s / m}} &
\textbf{\shortstack{of which\\refusals}} & \textbf{\shortstack{pass@1\\s / m}} &
\textbf{\shortstack{Emitted only\\s / m}} & \textbf{Cap} \\
\midrule
GPT-5.6-Terra       & 0 / 500 & 0 / 3,392 & 0 / 0 & 0 & 80.8 / 88.0 & 80.8 / 88.0 & 128k \\
GPT-5.6-Luna        & 0 / 500 & 0 / 3,498 & 0 / 0 & 0 & 70.4 / 81.2 & 70.4 / 81.2 & 128k \\
Qwen3.8-27B         & 150 / 500 & 108 / 3,235 & 35 / 0 & 0 & 66.8 / 92.4 & 71.8 / 92.4 & 128k\textsuperscript{a} \\
Qwen3.8-27B (250k)  & 124 / 500 & --- & 21 / --- & 0 & 70.0 / --- & 73.1 / --- & 250k\textsuperscript{a} \\
Qwen3.8-Flash-Next  & 51 / 500 & 92 / 3,141 & 20 / 0 & 0 & 84.2 / 93.0 & 87.7 / 93.0 & 128k \\
DeepSeek-V4.1-Flash & 11 / 500 & 49 / 3,525 & 10 / 1 & 0 & 88.4 / 92.0 & 90.2 / 92.2 & 128k \\
Claude Fable 5      & 3 / 500 & --- & 9 / --- & 6 & 90.4 / --- & 92.1 / --- & 128k \\
\bottomrule
\end{tabular}
\vspace{0.6ex}
\begin{minipage}{0.92\linewidth}
\footnotesize
\textsuperscript{a}~ Qwen3.8-27B's single arm was run at 250k and cap-matched back to 128k, where a cap hit is a completion reaching 128,000 output tokens.
\end{minipage}
\end{table}

Cap hits are rare across all models except the Qwen3.8 family, especially Qwen3.8-27B. Many runs that reached the cutoff still contained a complete solution earlier in their output or reasoning.
Of the single arm's 35 empty cells, the manager
solves 28, fails 7, and leaves none empty, worth 5.6 points. To investigate the effect of the cap, Qwen3.8-27B's
single arm was allowed to generate up to 250k tokens. This higher limit raised the score by 3.2
points, to 70.0. Re-scoring each single arm over only the problem-passes
where it emitted code bounds the effect model by model: the \emph{emitted only} column of
Table~\ref{tab:caps} carries that read for every arm, and puts Qwen3.8-27B's single call at 71.8.
Fable 5
gains 1.7 points there, six of its nine empty cells being safety refusals rather than
truncations.

\subsection{The OpenRouter-served model set}
\label{sec:openrouter}

The five models below were served through the OpenRouter gateway and run on the original
scaffold, so they are not directly comparable to the pinned-backend arms above.
Table~\ref{tab:openrouter} collects every condition.

\paragraph{Reasoning on:} Opus-5 scores at $88$ single $\to 95$ manager, the highest of all tested scenarios. However, smaller models tend to gain more from the manager (Figure~\ref{fig:scale-bars}a).

\paragraph{Reasoning off.} We ran five independent
passes per condition at a 16k cap plus a one-pass 128k control. Run-to-run spread across the
five passes is small. Kimi-K3 exhibits a particularly large gain, $+31.6$ points
(Figure~\ref{fig:scale-bars}b; Figure~\ref{fig:scale-16k} by scale). The 128k control largely
eliminates truncation and reproduces the ordering (panel c; Figure~\ref{fig:scale-off}).

\section{Discussion}
\label{sec:discussion}

Table~\ref{tab:transcripts} collects one transcript per model in which the single call failed
and the manager arm solved the problem. Three mechanisms recur. The scaffold commits the
reduction or the bounding argument to disk before any code is written, so the next call builds
on it instead of re-deriving it. It splits coupled objectives into pieces that are implemented
separately. And it bounds each call, so a model that would otherwise talk past its output cap
gets its work written down. In two cases the manager's solution is also markedly smaller than
the failed single-call one.

Truncation is a second place the scaffold earns its keep. State lives in \texttt{plan.md},
\texttt{notes.md} and \texttt{solution.py} rather than in one ever-growing transcript, and
each worker sees a curated view of it - plan, notes, current solution, one task - instead
of the full history. The workspace is external memory that the token cap cannot truncate.

That matters because a generation left to run to a very large cap does not merely risk being
cut off; it can degrade the model. Eight of the Qwen3.8-27B single calls became unstable during reasoning and
used up the full 250k budget without emitting code - on \texttt{abc399\_e},
the model repeated the phrase "\texttt{Alphabet 5 with a<->b, c<->d, e<->e fixed, and one letter? no.}" 7{,}743 times.
This is the mechanism behind the otherwise odd finding of Section~\ref{sec:truncation} that most
cut-off generations still contain a complete solution: the model had reached an answer and
then talked past the cap re-checking it. Bounding each call keeps the work far from the cap and
gets it onto disk.

The same argument should extend beyond this benchmark, though we do not test it. A task whose
inputs or intended output do not fit in one call - a codebase larger than the context window,
a deliverable longer than the output ceiling - is not solvable in one call at any reasoning
effort, whereas a sequence of scoped tasks can make progress.

The manager helps most where the base model is least able to organize its own reasoning, and
that weakness takes two forms. The first is having no internal planning stage at all: with
reasoning disabled a single call jumps straight to code, and the scaffold supplies a planning
space, on disk and across calls, that the model would not otherwise use
(Figure~\ref{fig:scale-bars}c). The second is reasoning that runs away: Qwen3.6-35B with
reasoning on spends its whole budget deliberating, and gains more here than in any of its other
conditions (panel a). Where a model can both plan internally and stop on
its own, the scaffold has less to add.

The manager is not free: its deliberation can settle on a worse answer than a single call would
have produced. On LCB 3765 (Qwen3.6-35B, 128k, reasoning off) the ideation stage identified the
same optimization the single call had used, talked itself out of it as too error-prone, and
committed instead to an asymptotically slower plan that a worker then implemented with a further
bug - compact, correct code disturbed rather than helped by decomposition. Qwen3.6-35B with
reasoning off is the standout loser on this pattern, negative at both caps.

\textbf{Two inexpensive, training-free additions could address the two halves of that failure},
the plan that was wrong and the implementation that went unchecked. First,
fresh-perspective workers: every worker inherits the notes and the current solution, so a
wrong early approach anchors everything downstream. Spawning some workers with the raw problem
statement and no prior context would give the manager an independent attempt to compare against,
protecting the tight single-call solution the scaffold otherwise disturbs - the intuition
behind sampling-and-voting and debate \citep{li2024,du2023}. Second, verification instead
of trust: the manager currently accepts a worker's report on the basis of a sanity check against
the problem's own samples, so a confidently-wrong solution propagates unchecked and can overwrite
a correct intermediate answer. Running each candidate against a generated test suite before
acceptance would catch regressions of this kind.

\subsection{Limitations}
\label{sec:limitations}

\begin{itemize}
  \item Five models ran through OpenRouter's multi-provider routing. OpenRouter does
  not guarantee one quantization or build per model. Many providers also intermittently stalled, dropped the response
  mid-stream, returned error objects, cut off output below the requested cap, or returned
  reasoning-only replies, resulting in a time-varying confound.

  \item Only one pass was run for the reasoning-on and 128k
  reasoning-off comparisons to limit API cost, so their results are a point estimate only.

  \item Fable 5 was run single-only due to its high cost, so it is unknown how much improvement its manager arm would gain.

  \item Fable 5
  returned \texttt{stop\_reason=refusal} on 6 of 500 problem-passes, resulting in a slight score reduction of 0-1.2 points.

  \item All results are based on code
  generation. Math and knowledge benchmarks (AIME, MATH-500, GPQA, HLE) were explored but abandoned due to saturation at near perfect scores.
\end{itemize}

\section{Conclusion}
\label{sec:conclusion}

The GVS5H scaffold improves the coding accuracy of each pinned-backend model run in both arms. These results provide a baseline against which more complex systems involving heterogeneous models or code verification can be compared.
The scaffold, despite increasing token use and hence cost compared to a model's single call, often remains
cheaper than frontier models while providing a higher accuracy. Transcript analysis
attributes the gain to decomposition and to context that persists across calls.   The result is conditional - some configurations result in no improvement or regression. However, in many cases, the orchestration is a cost-effective substitute for model scale.

\subsection*{AI use statement}

In this work, we used generative artificial intelligence as the object of study: every experimental result is a
measurement of language-model output, produced by the scaffold and baselines described in
Section~\ref{sec:method} and released with the code. Separately, generative language-model based tools assisted
with reviewing the manuscript, with writing the AI use, Ethics, and Reproducibility Statements, with implementing the orchestration harness and the
analysis and plotting scripts, and with literature search. We did not use generative artificial intelligence tools to
generate, impute or alter any experimental result, and all reported numbers are computed by the
released scripts directly from the stored run outputs. We have reviewed all artificial intelligence-assisted work: the
harness and analysis code was read and tested by the authors, the corrected evaluator of
Appendix~\ref{app:evaluator} was found and verified by manual inspection of transcripts, and every
citation was checked against its primary source. We take responsibility for the final content of
this work, including text, claims and artifacts produced with the aid of generative artificial intelligence.

\subsection*{Ethics statement}

This work involves no human subjects, no personal data and no new dataset release; the evaluation
set is the public hard split of LiveCodeBench release\_v6 \citep{jain2024}. Two points bear
on research integrity. First, one arm returned safety refusals that our protocol scores as
failures (Section~\ref{sec:limitations}); we report the rate rather than substituting a fallback
model, so a cell labeled with a model contains that model's outcome. Second, we report two defects in
a widely used public evaluation harness and its answer keys (Appendix~\ref{app:evaluator}) and the
corrected re-scoring of every number, rather than only the numbers that favor our method. The method itself
is a general-purpose inference-time scaffold for code generation and inherits the dual-use profile
of the underlying models; it lowers the cost of capable code generation, which applies to
beneficial and to harmful code alike.

\subsection*{Reproducibility statement}

Section~\ref{sec:method} specifies the manager--worker loop and its guards;
Section~\ref{sec:setup} and Appendix~\ref{app:protocol} give
the benchmark selection, the per-role temperatures, the serving path and provider settings per model, the output caps and
pass counts, the instrumentation, and the cap-matching replay procedure for Qwen3.8-27B's single
arm. Appendix~\ref{app:evaluator} documents both evaluator defects and their fixes, without which the
reported numbers cannot be reproduced from the stored generations. The statistical procedures are
given there too. Two sources of irreducibility are stated in
Section~\ref{sec:limitations}: decoding temperature is non-zero, and the OpenRouter-served arms
depend on provider health at run time. The pinned-backend arms, which carry the paper's claims,
avoid the second.

\subsection*{Code availability}

\iffinalversion
The orchestration scaffold (both versions), the corrected LiveCodeBench evaluator, and the
analysis and figure scripts that produce every number and figure in this paper are available at
\url{https://github.com/slee-persis/GVS5H}, together with all run outputs supporting the results
- per-problem generations for every arm and condition, per-call \texttt{finish\_reason} and
token logs where instrumented, and the manager arms' workspace transcripts (\texttt{runs/}).
\fi

\iffinalversion

\fi

\iffinalversion
%
%
%
\fi

\bibliography{references}
\bibliographystyle{paper_conference}

\appendix

\section{Benchmark and protocol in full}
\label{app:protocol}

\paragraph{Caps.} The 128k figure is a per-call \texttt{max\_tokens} bound, not a context-window
limit; for Opus-5, Fable 5 and the GPT-5.6 family it is also the provider's hard ceiling on one
response \citep{anthropic2026,openai2026}. Locally served Qwen3.8-27B has no such ceiling; it is
bounded by its 262{,}144-token context window \citep{qwen2026c} shared by prompt and output,
which puts the 250k setting of Section~\ref{sec:truncation} near its largest possible output.

\paragraph{Conditions.} The pinned-backend set runs at one setting: 128k output cap, reasoning on,
$\times 5$ passes, with the five models in both arms and Fable 5 single-only. The
OpenRouter-served set runs at three: 128k reasoning-on $\times 1$; 16k reasoning-off
$\times 5$, whose repeated passes give run-to-run variance and paired tests; and 128k reasoning-off
$\times 1$, the same without the small cap, isolating it as a confound.

\paragraph{Reasoning.} The output cap and the scaffold are matched across the eleven pinned-backend
arms; reasoning depth is not, because no two of these providers expose the same control over it.
Qwen3.8-27B, on our own vLLM, is sent no \texttt{reasoning\_effort}. The Qwen chat template
leaves reasoning on and unbudgeted, bounded only by the 128k \texttt{max\_tokens}, which on this
stack covers reasoning and answer together. The two GPT-5.6 models are likewise sent none, leaving
them at OpenAI's model default. Fable 5 runs at \texttt{thinking: adaptive} with \texttt{display:
summarized}, and \texttt{output\_config.effort} at \texttt{high}, the API default
\citep{anthropic2026effort}.

\paragraph{Instrumentation.} The two 128k-cap conditions log every call's \texttt{finish\_reason}
and completion-token count, and give each problem a final status in \{\texttt{ok},
\texttt{truncated}, \texttt{empty\_stop}, \texttt{empty}, \texttt{error}\}. The 16k $\times$ 5-pass
runs predate this and record only the extracted code and its pass/fail. An empty code field scores
as a failure.

\paragraph{Cap-matching.} One arm is not natively cap-matched to its partner: Qwen3.8-27B's single
call was generated at 250k while its manager arm ran at 128k (Section~\ref{sec:truncation}). To
compare them at one cap we replay the stored generations rather than re-run the model: each is
truncated to 128,000 output tokens, the solution re-extracted from that prefix, and the result
scored on the same evaluator as every other cell (Appendix~\ref{app:evaluator}): 

\begin{itemize}
  \item \textbf{Truncation is token-exact}, using the serving stack's own tokenizer instead of a
  character-count approximation.

  \item \textbf{What is truncated is the whole output stream, reasoning then answer}, because that
  is what the original cap bounded. Where the cut lands decides what the extractor sees: past the
  reasoning it sees a truncated answer; inside the reasoning the answer does not exist at all, and
  the harness's empty-content fallback hands it the truncated reasoning instead. (Section~\ref{sec:truncation}).

  \item \textbf{The cap is invisible to the model.} This arm
  ran on our own vLLM, where \texttt{max\_tokens} is enforced by the server: it never enters the
  prompt, and generation simply cuts off at the limit, abruptly rather than winding up (124 of the
  500 calls end at \texttt{finish\_reason=length}). Nor is this local to vLLM: making a
  budget visible to a model takes a separate, opt-in mechanism such as Anthropic's
  \texttt{task\_budget}, while \texttt{max\_tokens} stays an enforced ceiling that truncates the
  response \citep{anthropic2026effort}. The replay therefore does not differ from a genuine re-run
  at 128k.
\end{itemize}

\section{Two corrections to the LiveCodeBench evaluator}
\label{app:evaluator}

While inspecting transcripts for Section~\ref{sec:discussion} we found two defects in the grading
path: one in the LiveCodeBench harness, one in its answer keys. Every number in
Section~\ref{sec:results} is reported after fixing both; no model was re-run.
Table~\ref{tab:regrade} gives every reported arm as the stock evaluator graded it and after each
correction. Neither correction is a constant offset, and the larger of the two is the answer keys
for some models and the harness for others.

\paragraph{The \texttt{sys.stdin} mock.} The evaluator does not run a candidate as a subprocess.
Instead, it executes it in-process with
\texttt{sys.stdin} replaced by a mock. That mock's binary view implemented \texttt{readline()} as
\texttt{inputs.split(b"\textbackslash n")[0]}, which is a stateless expression that returns the
first line on every call. A program reading multi-line input through
\texttt{sys.stdin.buffer.readline()} therefore read line 1 repeatedly and failed however correct it
was. The mock's text-mode \texttt{sys.stdin.readline} was implemented as a proper iterator and
\texttt{buffer.read()} returns the whole payload, so neither was affected, which is why the
common \texttt{sys.stdin.buffer.read().split()} idiom never exposed it. We replaced the mock's
binary view with one backed by a \texttt{BytesIO} so that reads advance a position, and re-scored
every stored generation.

We report it explicitly for two reasons. First, it interacts badly with the v2 verifier: that step
runs the candidate as a real subprocess (step 5 of Section~\ref{sec:method}), where
\texttt{buffer.readline()} works correctly, so the manager was told its solution passed the public
samples for programs the grader then marked wrong - the loop's one external signal certifying a
wrong answer. Second, the exposure is uneven across models rather than a constant offset: 371 of the
5{,}425 pinned-backend outputs containing code use the idiom, and 86\% of those were scored wrong
against 18\% for the outputs that do not. Fable 5 never uses it and its scores are unchanged to the
decimal. The OpenRouter-served set reaches for it once in 4{,}764 generations containing code, and
then through an alias - Kimi-K3's single call on \texttt{abc396\_g} binds
\texttt{sys.stdin.buffer} to a name and calls \texttt{readline()} on that. Every OpenRouter-served
number is re-scored all the same, and the repaired mock moves one cell: Kimi-K3's 128k
reasoning-on single arm, from 82 to 83 (Table~\ref{tab:regrade}, where that cell reaches its
reported 86 through the answer-key correction below). Because the idiom is a matter of coding
style, a harness bug of this shape is not a wash across a leaderboard - it penalizes the models
whose style trips it, and its $\Delta$ harness column ranges from $+0.0$ for Fable 5, which
never writes it, to $+12.6$ for GPT-5.6-Terra's manager arm.

\paragraph{Answer keys that accept one of many valid answers.} LiveCodeBench grades by exact
string match against a single stored output. Four LCB-100 problems accept more than one:
\texttt{abc396\_e}, \texttt{arc190\_a} and \texttt{arc195\_c} ask for any construction that
attains the optimum, and \texttt{abc385\_f} asks for a real number, which its contest accepts
within a relative $10^{-6}$. We judge those four by the rule their contest used. Each of the three
constructive problems gets a checker that parses the submitted answer, tests it against the
problem's own constraints, and requires it to attain the objective value of the stored reference:
for \texttt{abc396\_e}, that the sequence is non-negative, satisfies every XOR constraint and sums
to what the reference sums to; for \texttt{arc190\_a}, that the chosen operations cover $1..N$ and
number what the reference says is optimal; for \texttt{arc195\_c}, that the Yes/No verdict matches
the reference and that any placement offered uses the requested counts of each colour, on distinct
cells, with every consecutive pair a step of the kind its colour allows. \texttt{abc385\_f} is
compared token by token at $10^{-6}$ relative tolerance. This class of defect is one LiveCodeBench documents upstream - its errata
list eleven earlier problems whose keys accept multiple answers, none of them in LCB-100 - and
it is a property of the benchmark, not of any model, so it applies to every arm of both sets and
moves 104 of their 118 arm-passes. Re-judging
runs each submission as a real subprocess rather than in-process, which is also what repairs the
mock above and makes \texttt{sys.stdout.buffer} write as it does on the contest judge.

Both corrections are conservative in the same two ways. They re-run only the submissions the stock
evaluator marked wrong, so neither can turn a pass into a failure and the corrections move every
arm of Table~\ref{tab:regrade} upward; and among those they re-run only
the ones with a known reason to be misjudged - a failure on one of the four problems above, or
code that touches \texttt{sys.stdin.buffer} or \texttt{sys.stdout.buffer}. A submission that the
in-process evaluator failed for some other reason keeps that verdict. Re-executing every stored
generation instead, passing and failing alike, finds 25 further submissions across 11{,}800
records that pass every hidden test as a subprocess; they are spread over 22 arms, at most three
to an arm, and we leave them as the stock evaluator scored them.

\begin{table}[htbp]
\centering
\caption{\textbf{Every reported arm as the stock evaluator graded it, and after each
correction} (Appendix~\ref{app:evaluator}). \textbf{As graded} is LiveCodeBench's own
verdict on the stored generations. \textbf{Harness} is the move contributed by running
each submission as a subprocess, which is what repairs the \texttt{sys.stdin} mock and
\texttt{sys.stdout.buffer}; \textbf{keys} is the move contributed by judging the four
problems that accept more than one answer by the contest's rule. \textbf{Reported} is the
number used everywhere else in this paper. Upper block: five-pass means at a 128k cap,
reasoning on (Table~\ref{tab:main}). Lower block: the 128k reasoning-on, one-pass
OpenRouter arms (Table~\ref{tab:openrouter}); Qwen3.5-9B is unusable in this condition and
is omitted. Across the 48 arm-passes of the two remaining OpenRouter conditions the harness
correction moves six by one point and the key correction moves each arm by 0 to 3 points.
Qwen3.8-27B's single call is the one row where \textbf{as graded} is not the stock evaluator
on the code being reported: that arm is cap-matched by replaying its 250k generations at 128k
(Appendix~\ref{app:protocol}) and the truncated solutions were never put through the stock
evaluator, so the column is the 250k grading and the two $\Delta$ columns carry the cap match
as well as the corrections. Against that baseline the row gains 35 problem-passes and loses
13, eight of them generations whose 128k prefix contains no code at all.}
\label{tab:regrade}
\small
\setlength{\tabcolsep}{4pt}
\begin{tabular}{@{}lcccc@{\hspace{1.2em}}cccc@{}}
\toprule
& \multicolumn{4}{c}{\textbf{Single call}} & \multicolumn{4}{c}{\textbf{With manager}} \\
\cmidrule(r){2-5}\cmidrule(l){6-9}
\textbf{Model} & \textbf{\shortstack{As\\graded}} & \textbf{\shortstack{$\Delta$\\harness}}
& \textbf{\shortstack{$\Delta$\\keys}} & \textbf{\shortstack{Re-\\ported}}
& \textbf{\shortstack{As\\graded}} & \textbf{\shortstack{$\Delta$\\harness}}
& \textbf{\shortstack{$\Delta$\\keys}} & \textbf{\shortstack{Re-\\ported}} \\
\midrule
\multicolumn{9}{@{}l}{\emph{Pinned backends, five passes}} \\
Claude Fable 5 & 87.4 & $+0.0$ & $+3.0$ & \textbf{90.4} & --- & --- & --- & --- \\
GPT-5.6-Terra & 69.8 & $+8.4$ & $+2.6$ & \textbf{80.8} & 72.6 & $+12.6$ & $+2.8$ & \textbf{88.0} \\
GPT-5.6-Luna & 60.4 & $+8.0$ & $+2.0$ & \textbf{70.4} & 70.4 & $+8.4$ & $+2.4$ & \textbf{81.2} \\
Qwen3.8-27B & 62.4 & $+2.6$ & $+1.8$ & \textbf{66.8} & 84.4 & $+5.4$ & $+2.6$ & \textbf{92.4} \\
Qwen3.8-Flash-Next & 78.8 & $+2.4$ & $+3.0$ & \textbf{84.2} & 87.4 & $+2.6$ & $+3.0$ & \textbf{93.0} \\
DeepSeek-V4.1-Flash & 84.2 & $+1.2$ & $+3.0$ & \textbf{88.4} & 87.6 & $+1.4$ & $+3.0$ & \textbf{92.0} \\
\midrule
\multicolumn{9}{@{}l}{\emph{OpenRouter-served, one pass}} \\
Opus-5 & 85.0 & $+0.0$ & $+3.0$ & \textbf{88.0} & 91.0 & $+1.0$ & $+3.0$ & \textbf{95.0} \\
Kimi-K3 & 82.0 & $+1.0$ & $+3.0$ & \textbf{86.0} & 82.0 & $+1.0$ & $+3.0$ & \textbf{86.0} \\
MiniMax-M3 & 60.0 & $+0.0$ & $+2.0$ & \textbf{62.0} & 66.0 & $+0.0$ & $+3.0$ & \textbf{69.0} \\
Qwen3.6-35B & 25.0 & $+0.0$ & $+1.0$ & \textbf{26.0} & 43.0 & $+0.0$ & $+1.0$ & \textbf{44.0} \\
\bottomrule
\end{tabular}
\end{table}

\section{Additional results}
\label{app:results}

\begin{chartfigure}
\centering
\halfplot{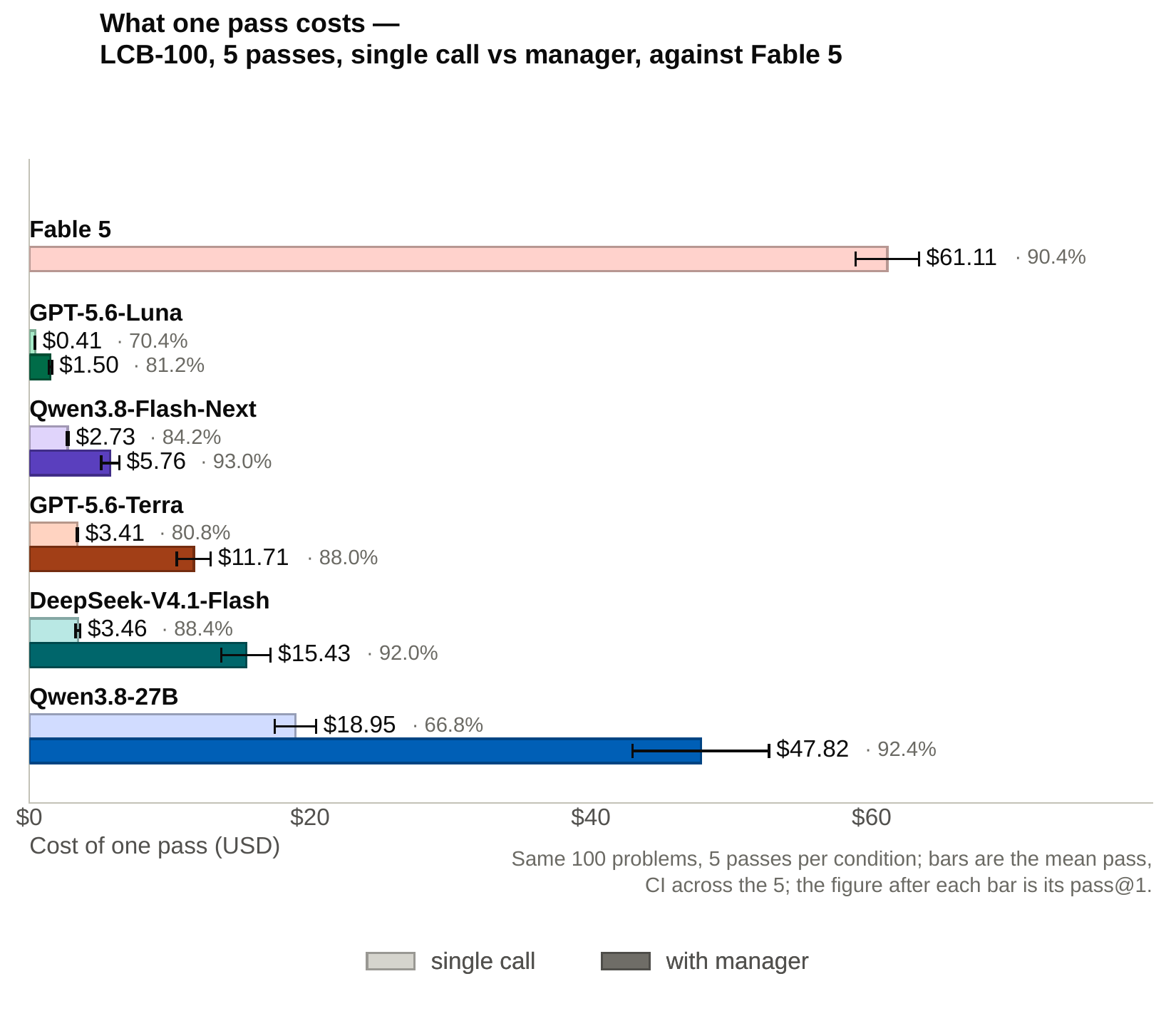}
\caption{\textbf{The scaffold's bill.} Cost of one pass over the same 100 problems; Table~\ref{tab:cost} is the arithmetic behind every bar. No cached-input discount is taken, and Qwen3.8-27B is priced at OpenRouter market rates. The top bar of each pair is the single call, light; the manager is under it, dark; all are at a 128k output cap, and colour is the model. Fable 5 is first; the models below it run top to bottom from the cheapest single call to the dearest. Table~\ref{tab:cost} carries the tests. Unbracketed comparisons: every manager $-$ single increase is significant, and GPT-5.6-Luna's manager undercuts GPT-5.6-Terra's single call by \$1.91. At 1.28x the assumed Qwen rate (\$0.273/\$3.26 per MTok, inside the spread across hosted providers) the manager's cost advantage over Fable 5 disappears entirely — a larger uncertainty than any p-value here.}
\label{fig:cost}
\end{chartfigure}

\begin{table}[htbp]
\centering
\caption{\textbf{Problem-passes won by one arm alone.} Of the 500 problem-passes per model (100 problems $\times$ 5 passes), the number the manager arm solved and the single call did not, against the number the single call solved and the manager did not. The gains are not compensating wins and losses. The test is exact McNemar on those discordant pairs (Section~\ref{sec:setup}).}
\label{tab:agreement}
\small
\begin{tabular}{@{}lcccc@{}}
\toprule
\textbf{Model} & \textbf{Manager only} & \textbf{Single call only} &
\textbf{Problem-passes} & \textbf{Exact McNemar $p$} \\
\midrule
GPT-5.6-Terra       & 47 & 11 & 500 & $2.0\times10^{-6}$ \\
GPT-5.6-Luna        & 72 & 18 & 500 & $8.1\times10^{-9}$ \\
Qwen3.8-27B         & 134 & 6 & 500 & $1.4\times10^{-32}$ \\
Qwen3.8-Flash-Next  & 49 & 5 & 500 & $3.9\times10^{-10}$ \\
DeepSeek-V4.1-Flash & 22 & 4 & 500 & $5.3\times10^{-4}$ \\
\bottomrule
\end{tabular}
\end{table}

\begin{table}[htbp]
\centering
\caption{\textbf{LCB-100 pass@1 (\%), single $\to$ manager.} For the OpenRouter-served models on the
original scaffold. ``--'' = not run; Qwen3.5-9B reasoning-on returns reasoning-only replies and is
unusable. Scores are from the corrected grader (Appendix~\ref{app:evaluator}): the four LCB-100 problems
whose reference answer is not unique are judged by the original contest rule, and submissions run
as real subprocesses so \texttt{sys.stdout.buffer} behaves as it does on the contest judge. A
bolded delta clears the ${\sim}4.5$ point pass-to-pass band. The eleven pinned-backend, five-pass
arms are reported separately in Section~\ref{sec:results} and are not pooled here, because both the
serving path and the scaffold version differ (Section~\ref{sec:setup}).}
\label{tab:openrouter}
\small
\begin{tabular}{@{}llccc@{}}
\toprule
\textbf{Model} & \textbf{Params} & \textbf{128k $\cdot$ ON (1 pass)} &
\textbf{128k $\cdot$ OFF (1 pass)} &
\textbf{16k $\cdot$ OFF ($\times$5)} \\
\midrule
Opus-5      & n/a            & 88 $\to$ 95 (\textbf{+7}) & -- & -- \\
Kimi-K3     & ${\sim}2.8$T   & 86 $\to$ 86 (+0) & 33 $\to$ 76 (\textbf{+43}) & 33.4 $\to$ 65.0 (\textbf{+31.6}) \\
MiniMax-M3  & 428B           & 62 $\to$ 69 (\textbf{+7}) & 26 $\to$ 40 (\textbf{+14}) & 22.2 $\to$ 33.2 (\textbf{+11.0}) \\
Qwen3.6-35B & 35B            & 26 $\to$ 44 (\textbf{+18}) & 35 $\to$ 27 ($-$8) & 28.2 $\to$ 27.4 ($-$0.8) \\
Qwen3.5-9B  & 9B             & \emph{unusable} & 18 $\to$ 21 (+3) & 15.0 $\to$ 22.6 (\textbf{+7.6}) \\
\bottomrule
\end{tabular}
\end{table}

\begin{table}[p]
\centering
\caption{\textbf{One transcript per model where the single call failed and the manager arm solved
the problem} (Section~\ref{sec:discussion}). LCB problem IDs as in release\_v6; call counts are
manager-arm model calls. Character counts are of the emitted program.}
\label{tab:transcripts}
\footnotesize
\setlength{\tabcolsep}{4pt}
\setlength{\defaultaddspace}{2.5pt}
\begin{tabular}{@{}p{0.155\linewidth}p{0.101\linewidth}p{0.310\linewidth}p{0.364\linewidth}@{}}
\toprule
\textbf{Model} \newline \textbf{(condition)} & \textbf{Problem} & \textbf{How the single call failed} &
\textbf{What the scaffold added} \\
\midrule
Qwen3.5-9B \newline (128k, off) & abc385\_d & Timed out: a na\"ive per-segment scan of a path
simulation. & The brainstorm flagged in writing that the check is $O(N\cdot M) \approx
4\times10^{10}$ and prescribed a range-query structure before any code - the up-front
planning a single pass skips (Section~\ref{sec:discussion}). \\
\addlinespace
Qwen3.6-35B \newline (on) & abc394\_f & Cut off mid-reasoning at 32,768 tokens; returned nothing. &
Solved in one worker cycle: the brainstorm fixed the structural insight (an ``alkane'' subtree needs
a degree-4 center and $\ge 5$ vertices); the worker used an iterative DFS to avoid a recursion-depth
failure. The same 32k clamp was hit 117 times across this model's manager calls, but with several
attempts per problem one clamped call no longer costs the problem. \\
\addlinespace
MiniMax-M3 \newline (on) & 3701 & Printed nothing on \texttt{cdcd}: the lexicographic-reconstruction
branch was broken. & Split the two coupled objectives across worker cycles - one worker on the
forward cost DP over capped run-lengths, a separate worker on a suffix DP used purely for the
lexicographic reconstruction. \\
\addlinespace
Kimi-K3 \newline (128k, off) & 3687 & Over-counted the longest unique-value path (9 and 3 where the
answers were 6 and 2), never shrinking its window on a repeated value. & The brainstorm wrote the
reduction into \texttt{notes.md} first - longest substring without repeating characters along each
root-to-leaf path, window start via last-seen depths, $O(n)$ DFS. Only then did a worker
implement it. \\
\addlinespace
GPT-5.6-Terra \newline (on) & 3688 & 4,615 characters implementing a Segment-Tree-Beats-shaped
structure indexed by candidate value, and wrong. & The plan named the difficulty in advance - the
state must be compressed rather than maintaining all distinct values per index. Nine calls
later, the manager arm returned 1,860 characters over one standard segment tree. The contribution is
\emph{subtraction}. Manager-only win in four of five passes. \\
\addlinespace
GPT-5.6-Luna \newline (on) & abc397\_e & Carried a \emph{set} of unfinished path lengths per subtree
across 4,092 characters, a general state it never closed. & The brainstorm established the bounding
lemma (at most two unfinished paths at any vertex), after which a worker implemented the postorder
scan directly in 2,166 characters over five calls. Manager-only win in two of five passes. \\
\addlinespace
Qwen3.8-27B \newline (on) & abc397\_g & Spent all 250,000 tokens on edge-case self-interrogation and
never emitted a program - the runaway-reasoning failure mode of Section~\ref{sec:discussion} in
pure form - though the reasoning had already reached the min-cut formulation. & Committed that
same reduction to \texttt{plan.md} as a plan, then had a worker implement it: five calls, 3,548
characters, manager-only win in all five passes. The scaffold does not supply the insight; it forces
it onto disk before the budget runs out. \\
\addlinespace
Opus-5 \newline (on) & abc388\_e & The feasibility test over-counted, reporting 225 pairs where 220
is optimal. & Recorded the justification in \texttt{notes.md} first - an exchange argument,
plus monotonicity of feasibility in $K$ licensing a binary search - then had one worker implement
it and a second harden it. \\
\bottomrule
\end{tabular}
\end{table}

\begin{figure}[p]
\centering
\begin{subfigure}[t]{0.32\linewidth}
  \centering
  \panelplot{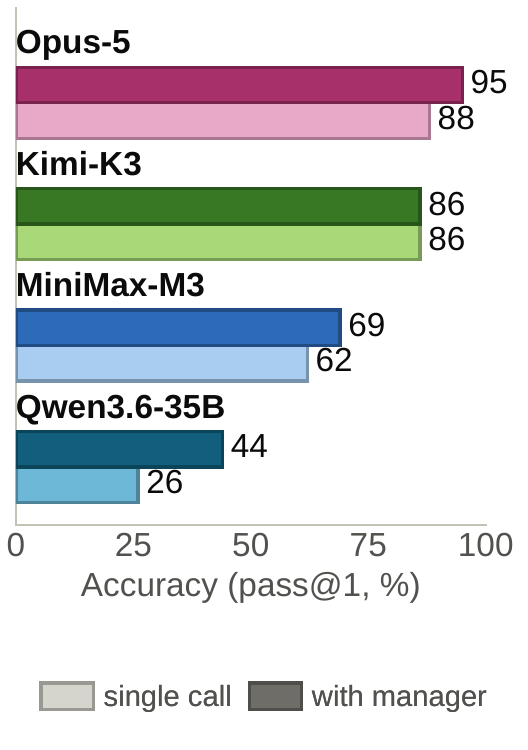}
  \caption{128k max tokens, reasoning on, one pass per condition.}
\end{subfigure}\hfill
\begin{subfigure}[t]{0.32\linewidth}
  \centering
  \panelplot{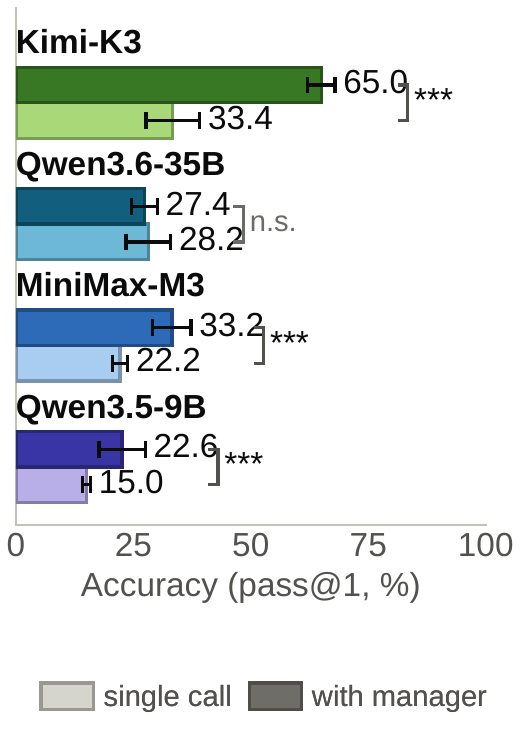}
  \caption{16k max tokens, reasoning off, five passes per condition.}
\end{subfigure}\hfill
\begin{subfigure}[t]{0.32\linewidth}
  \centering
  \panelplot{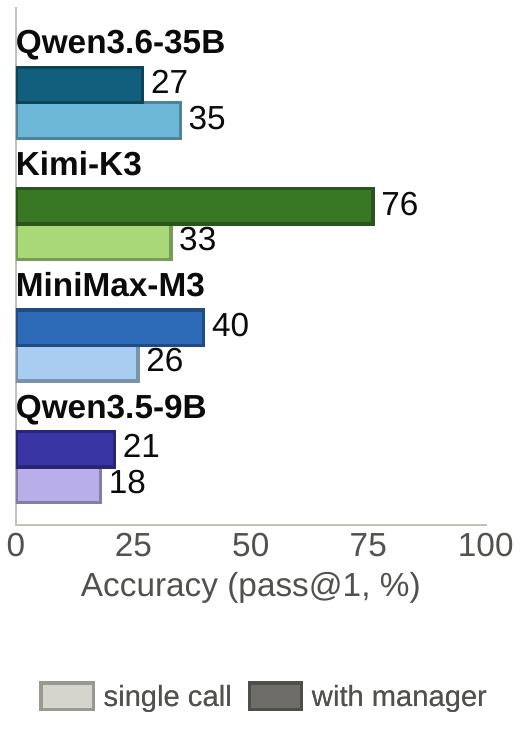}
  \caption{128k max tokens, reasoning off, one pass per condition.}
\end{subfigure}
\caption{\textbf{Manager vs.\ single call, three conditions.} Models run top to bottom by single-call score, best first, so a model's row differs between panels; the top bar of each pair is the manager. Fill: light = single call (one call, no tools), dark = with manager; $\Delta$ = manager $-$ single, in percentage points. The line through a bar is the 95\% CI across its passes, where the condition ran more than one. A panel prints its two bar values and nothing else; the $\Delta$ and Tukey HSD group letter quoted below per model are what it leaves out (shared letter = not significantly different across the models of that condition, $\alpha = 0.05$). (a) 128k max tokens, reasoning on - effort:high for Kimi-K3 and Opus-5, a 20k reasoning budget requested for Qwen3.6-35B and MiniMax-M3 that providers often did not honor. One pass per condition, so no per-model p. Non-empty completions per 100 (single$\to$manager): Opus-5 97$\to$100, Kimi-K3 97$\to$100, MiniMax-M3 92$\to$93, Qwen3.6-35B 34$\to$68. Empty or truncated output scores as a fail, so $\Delta$ partly tracks emit rate. Opus-5 $\Delta$+7.0 (group ab), Kimi-K3 $\Delta$+0.0 (group b), MiniMax-M3 $\Delta$+7.0 (group ab), Qwen3.6-35B $\Delta$+18.0 (group a). (b) 16k max tokens, reasoning off, five passes per condition; the line through each bar is the 95\% CI across the five passes (t, df = 4). Each $\Delta$ is a paired sign-flip permutation test, unit = problem (n = 100), Holm-corrected across the 4 models; the bracket beside a pair grades its p: * p $<$ .05, ** p $<$ .01, *** p $<$ .001, n.s. otherwise. Kimi-K3 $\Delta$+31.6 (group a), Qwen3.6-35B $\Delta$-0.8 (group c), MiniMax-M3 $\Delta$+11.0 (group b), Qwen3.5-9B $\Delta$+7.6 (group bc). (c) 128k max tokens, reasoning off (ESCALATION\_OR\_REASONING=none), one pass per condition, so no per-model p. Non-empty completions per 100 (single$\to$manager): Qwen3.6-35B 85$\to$95, Kimi-K3 96$\to$100, MiniMax-M3 93$\to$100, Qwen3.5-9B 79$\to$92. Qwen3.6-35B $\Delta$-8.0 (group c), Kimi-K3 $\Delta$+43.0 (group a), MiniMax-M3 $\Delta$+14.0 (group b), Qwen3.5-9B $\Delta$+3.0 (group bc).
}
\label{fig:scale-bars}
\end{figure}

\begin{figure}[htbp]
\centering
\panelplot{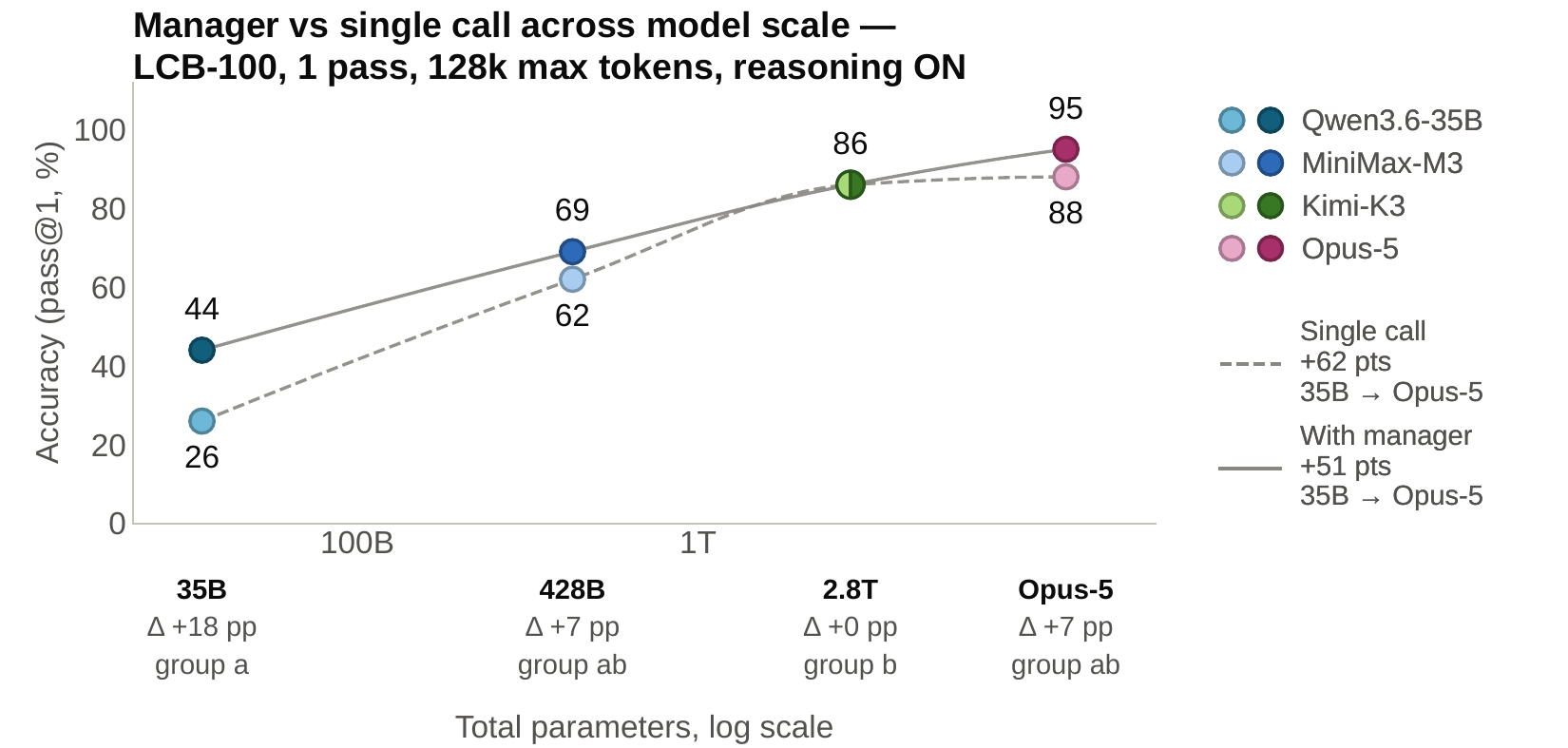}
\caption{\textbf{Accuracy by model scale, reasoning on, 128k, one pass.} The same numbers as Figure~\ref{fig:scale-bars}a, against parameter count. Dashed = single call, solid = with manager; the block under each tick gives that model's $\Delta$ and Tukey group, and the legend each arm's gain from the smallest model to the largest. }
\label{fig:scale-on}
\end{figure}

\begin{figure}[htbp]
\centering
\panelplot{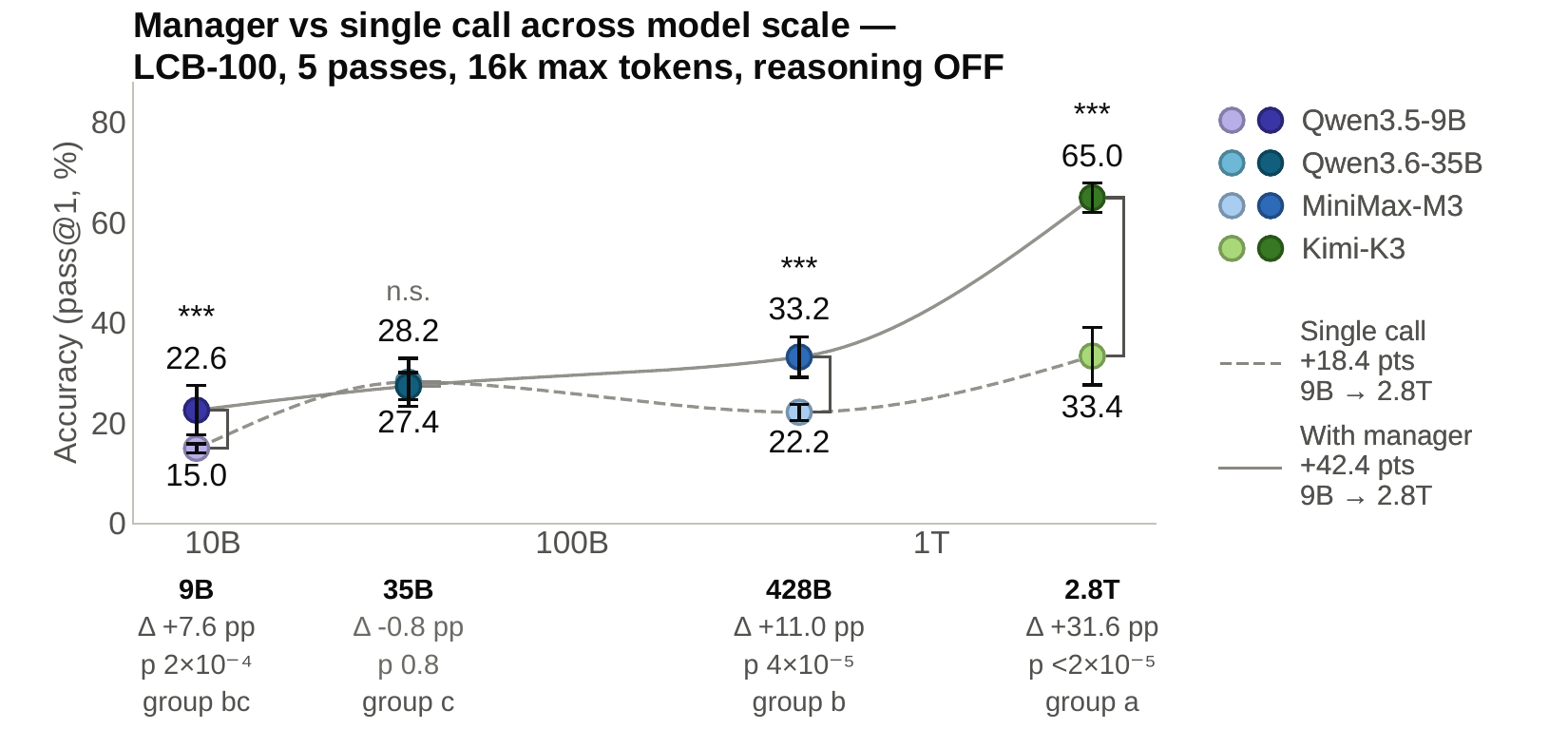}
\caption{\textbf{Accuracy by model scale, reasoning off, 16k, five passes.} The same numbers as Figure~\ref{fig:scale-bars}b, against parameter count; the line through each mark is the 95\% CI across the five passes. Dashed = single call, solid = with manager, and the bracket at a pair grades its Holm-corrected p; the block under each tick gives that model's $\Delta$, p and Tukey group, and the legend each arm's gain from 9B to 2.8T. }
\label{fig:scale-16k}
\end{figure}

\begin{figure}[htbp]
\centering
\panelplot{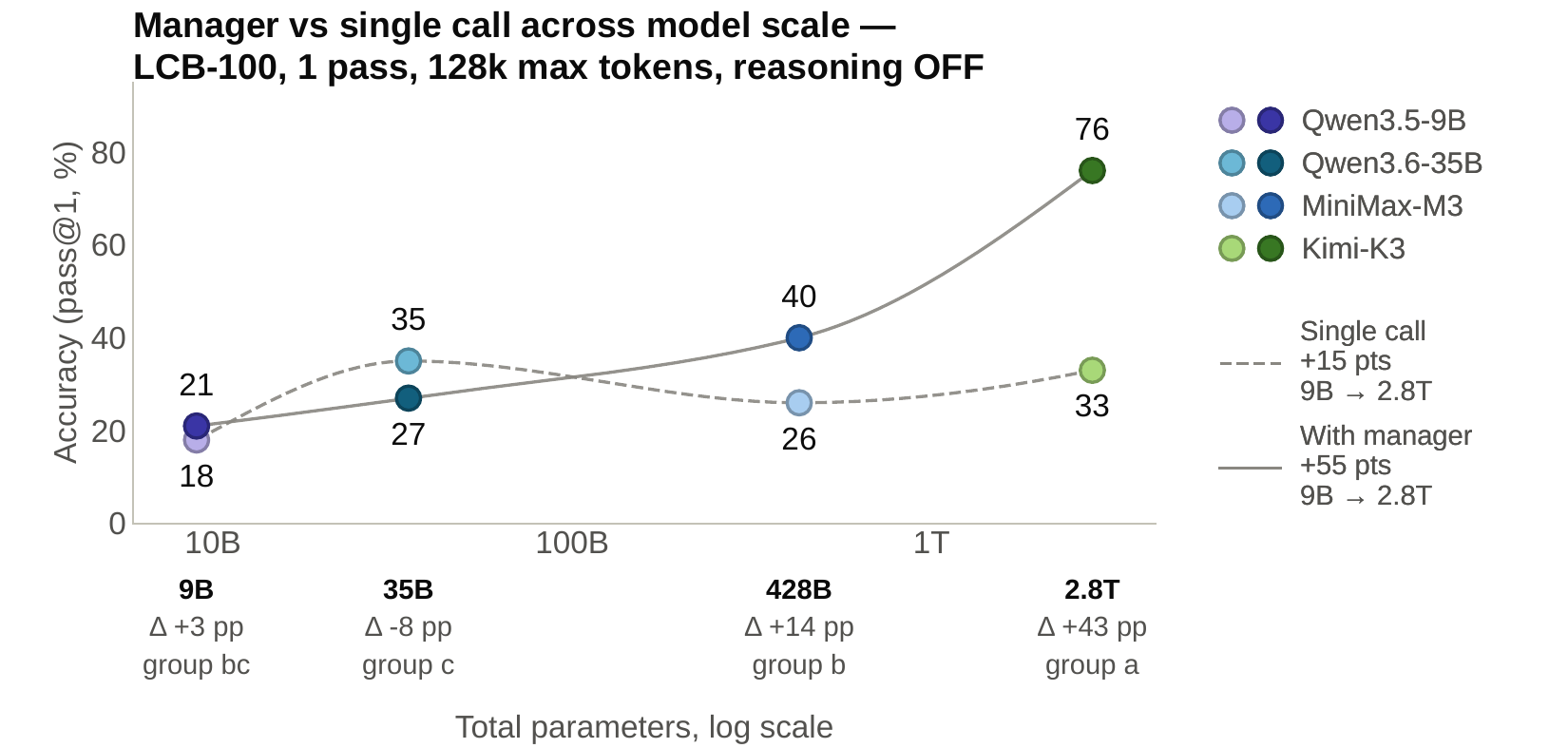}
\caption{\textbf{Accuracy by model scale, reasoning off, 128k, one-pass control.} The same numbers as Figure~\ref{fig:scale-bars}c, against parameter count. Dashed = single call, solid = with manager; the block under each tick gives that model's $\Delta$ and Tukey group, and the legend each arm's gain from the smallest model to the largest. }
\label{fig:scale-off}
\end{figure}

\end{document}